\documentclass[10pt]{iopart}
\usepackage{graphicx}
\begin{document}
\title{A new family of exactly solvable disordered reaction-diffusion systems}
\author{Mohammad Ghadermazi $^1$ and Farhad H. Jafarpour$^2$}
\address{$^1$Plasma Physics Research Center, Science and Research Branch, Islamic Azad University, Tehran, Iran}
\address{$^2$Physics Department, Bu-Ali Sina University, 65174-4161 Hamedan, Iran}
\begin{abstract}
Using a matrix product method the steady-state of a family of disordered reaction-diffusion systems consisting of different species of
interacting classical particles moving on a lattice with periodic boundary conditions is studied. A new generalized quadratic
algebra and its matrix representations is introduced. The steady-states of two members of this exactly solvable family of
systems are studied in detail.
\end{abstract}
\pacs{64.60.De, 82.40.Bj, 02.10.Yn}
\maketitle
\section{Introductions}

One-dimensional reaction-diffusion systems are systems of classical particles with stochastic dynamics and
hard-core exclusion~\cite{schutz}. The particle disorder version of a reaction-diffusion system is a
multi-species system of particles where the particle hopping rate of each species can be a quenched random variable. 
This means that the particles of different species have their own intrinsic rates for hopping forward and backward. 
On the other hand, in a site disorder version of a reaction-diffusion system the hopping  rate of a given particle depends 
on the site in which the particle resides. In this paper we are dealing with particle disorder examples.

Different types of exactly solvable multi-species reaction-diffusion systems have been introduced and studied in related 
literature~\cite{Benjamini,EvansEurophys,BasuMohanty1,Vahid,JafarpourRing}.
The particle disorder reaction-diffusion systems usually settle into a non-equilibrium steady-state in long-time limit. There are different 
approaches to study these non-equilibrium steady-states. The Matrix Product Formalism (MPF) is one of these techniques. 
According to this formalism one associates an operator to every state of a lattice-site of the system. It is assumed that the steady-state 
weight of each configuration of the system defined on a ring geometry is given by a trace of product of these non-commuting operators.
The basics of this technique can be summarized as follows: Let us consider a one-dimensional system, defined on a lattice with a ring geometry, 
in which each lattice-site can be in one of its $\nu+1$ accessible states. We define $\nu+1$ operators associated with different states of each lattice-site.
We associate the operator $D_{\tau_i} $ to the $i$'th lattice-site provided that it is occupied by a particle of type $\tau_i$ ($\tau_i=1,\cdots, \nu$).
If this lattice-site is empty $\tau_i=0$ then the operator $D_0 \equiv E$ will be associated with it. According to the MPF the steady-state weight 
of a given configuration similar to $100223040$ is proportional to $\Tr[D_1EED_2D_2D_3ED_4E]$. Requiring that these are the stationary 
weights provides us with an algebra for the operators $D_{\tau_i}$'s and $E$. For the systems with nearest-neighbors interactions this algebra is quadratic.

The MPF was first introduced in~\cite{DerridaEvansHakimPasquier} to find the steady-state of a single-species partially asymmetric 
exclusion process with nearest-neighbors interactions. It was then extended and generalized to find the steady-states of multi-species 
reaction-diffusion systems~\cite{IPR,Alcaraz1,Alcaraz2,Arndt,BlytheEvans} even with long-rage interactions~\cite{Andreas}. 

Finding the exact expressions for the steady-state weights using the MPF requires one to either work with the commutation relations of the operators
or find a matrix representation for the associated algebra of the system. To the best of our knowledge there is no straightforward approach to
predict whether a one-dimensional reaction-diffusion system has a matrix product steady-state. This makes finding exactly solvable
models with matrix product steady-states a formidable task.

The number of particles in each species of a one-dimensional multi-species reaction-diffusion system can be a conserved or a non-conserved quantity. 
As an example for a system with conserved number of particles the authors of~\cite{EvansEurophys} have studied 
a particle disorder driven-diffusive system consisting of $\nu$ different species of particles which hop in a preferred direction on a 
discrete ring with $L$ lattice-sites. It has been shown that the steady-state of this system can be obtained exactly using the MPF. It 
also turns out that this steady-state is a product measure. 

Exact matrix product solution for the stationary state measure of the partially asymmetric exclusion process on a ring with multiple species of particles
has also been obtained in~\cite{EFM,PEM}. This is also an example for a system with conserved number of particles in each species.

In another work~\cite{BasuMohanty1} the authors have introduced a totally asymmetric exclusion process on a discrete
ring with $\nu$ non-conserved internal degrees of freedom. In this process, apart from the usual forward hopping of particles,
the particle type can be changed i.e. a particle in one of $\nu$ possible internal states can be converted to any other $\nu-1$ states
with different rates. It has been shown that the steady-state of this system can be obtained using a matrix product approach. The
matrices associated with different states of a lattice-site satisfy an algebra which has a $\nu$-dimensional matrix representation.

So far most of the reaction-diffusion systems which have been studied using the MPF belong to the group of processes in which the number of particles in each species 
does not change with time~\cite{EvansEurophys,JafarpourRing,IPR,Alcaraz1,Alcaraz2,Arndt}. In this paper we introduce a new family of one-dimensional disordered reaction-diffusion systems whose steady-state can be obtained using the MPF; however, the number of particles in each species changes with time. 
As we will see the system already studied in~\cite{BasuMohanty1} is only a member of this family of exactly solvable systems.

We will show that the steady-state probability distribution of this family of exactly solvable systems is  factorized, similar to that of a Zero-Range Process (ZRP).
A ZRP is a model of interacting particles which hop between the sites of a lattice with rates that depend on the occupancy of the departure sites.
The interaction between the particles does not obey the exclusion principle in the sense that many indistinguishable particles might occupy a
single lattice-site. The ZRP has unique physical and mathematical properties~\cite{EvansHanney}. 

It is also known that some of these exclusion processes can be mapped onto the ZRP~\cite{EvansHanney,EvansBraz}.
We will show that our new family of exactly solvable systems can be mapped onto a ZRP. Similar systems have already been studied in related literature~\cite{BasuMohanty1,Jafarpour,BasuMohanty}.  We start with a disordered exclusion process defined on a one-dimensional lattice of finite size with 
periodic boundary conditions. Under some constraints on the microscopic reaction rates the steady-state weights can be calculated exactly using the MPF. 
Special cases are studied in more detail.

The paper is organized as follows: In section $2$ we introduce a generalized quadratic algebra and its matrix representations. The most general dynamical rules for the processes whose steady-states can be described by this quadratic algebra are presented. We have shown that under some specific constraints on the transition rates 
these steady-states are matrix product states. In section $3$ two special examples are studied in detail.

\section{A generalized quadratic algebra }

Let us consider a system consisting of $\nu$ different types of classical particles which hop on a discrete ring of $L$ lattice-sites. Each lattice-site is either occupied by one
type of particle or empty. On each lattice-site $i$ ($i=1,2,\cdots,L$) we allow for $\nu+1$ configurations described by means of the variable $\tau_i$, which takes $(\nu+1)$ values $0,1,2,\cdots,\nu$. For $\tau_i=0$ the lattice-site $i$ is empty and for $\tau_i=k$ the lattice-site $i$ is occupied by a particle of type $k$ where $k=1,2,\cdots,\nu$. For two adjacent lattice-sites $(i,i+1)$ the rate of exchange between the configuration ($\tau_i,\tau_{i+1}$) into the configuration ($\gamma_i,\gamma_{i+1}$) is represented by $\Gamma_{\gamma_i,\gamma_{i+1}}^{\tau_i,\tau_{i+1}}$. The time evolution of the probability distribution of a Markovian interacting particle system $|P(t)\rangle$ is given by a Schr\"{o}dinger like equation in imaginary time
\begin{equation}
\label{hamiltoni}
\frac{d}{dt}|P
(t)\rangle=H \vert P(t)\rangle
\end{equation}
in which $H$ is a stochastic Hamiltonian \cite{schutz}. The matrix elements of the Hamiltonian are the transition rates between different configurations.
For a one-dimensional system with nearest-neighbors interactions the Hamiltonian $H$ has the following general form
\begin{equation}
\label{general hamiltoni}
H=\sum_{i=1}^{L-1}h_{i,i+1}
\end{equation}
in which
\begin{equation}
\label{tansor}
h_{i,i+1}=I^{\otimes(i-1)}\otimes h \otimes I^{\otimes(L-i-1)}
\end{equation}
here $I$ is a $(\nu+1)\times(\nu+1)$ identity matrix and $h$ is a $(\nu+1)^2\times (\nu+1)^2$ matrix for the bulk interactions and its matrix elements are defied as
\begin{equation}
\label{hamiltonian}
\left(h_{i,i+1}\right)_{\gamma_i,\gamma_{i+1}}^{\tau_i,\tau_{i+1}}=\left\{\begin{array}{l}\left[\Gamma_{i,i+1}\right]_{\gamma_i,\gamma_{i+1}}^{\tau_i,\tau_{i+1}} \;\; \mbox{for\;\;$\gamma_j\neq\tau_j$ and $j=i,i+1$} \cr \cr
-{\sum'}_{\beta_i,\beta_{i+1}}\left[\Gamma_{i,i+1}\right]^{\tau_i,\tau_{i+1}}_{\beta_i,\beta_{i+1}} \;\; \mbox{else}
\end{array}\right.
\end{equation}
in which ${\sum'}$ denotes the sum where the term $(\tau_i,\tau_{i+1})=(\beta_i,\beta_{i+1})$ is excluded. The number of the particles of type $k$ is denoted by $N_{k}$
then the number of empty lattice-site is given by $N_0=L-\sum_{k=1}^{\nu}N_{k}$.

We are interested in the steady-state of the system i.e the right eigenvector of the Hamiltonian $H$ with eigenvalue zero. In what follows we will apply the MPF in order to find this eigenvector of $H$. As me mentioned above we associate the operator $D_{k}$ $(k=1,2,\cdots,\nu)$ to the presence of a particle of type $k$ at a given lattice-site. We also associate the operator $E$ to an empty lattice-site.

Let us assume that the operators $D_k$'s and $E$ satisfy the following generalized quadratic algebra
\begin{equation}
\label{Algebra2}
\begin{array}{lll}
D_{k} E &=& p_{k}D_{k}  \quad  k=1,\cdots,\nu, \\
D_{k} D_{k'} &=& q_{k}D_{k'} \quad  k,k'=1,\cdots,\nu
\end{array}
\end{equation}
in which the parameters $p_{k}$ and $q_{k}$ can be any non-zero positive or negative numbers. As we will see for any disordered reaction-diffusion system, these
parameters should be calculated  separately. We are looking for those disordered reaction-diffusion systems with the Hamiltonian of the form~(\ref{hamiltonian})
whose steady-states can be described by the quadratic algebra~(\ref{Algebra2}). One can easily verify that this generalized quadratic algebra has an
infinite-dimensional matrix representation given by the following matrices
\begin{equation}
\label{RepAlgebra}
D_{k}=q_k \sum_{i=0}^{\infty} p_k^i|0\rangle\langle i |\quad,\quad E=\sum_{i=0}^\infty |i+1\rangle\langle i |
\end{equation}
in which $\{|i\rangle\}$ is the standard basis of an infinite-dimensional vector space defined as
\begin{equation}
\label{base}
\vert i \rangle_j=\delta_{i,j}\;\;  \mbox{for}\;\; i,j=0,1,\cdots ,\infty\,.
\end{equation}
In addition to the matrix representation (\ref{RepAlgebra}), the quadratic algebra (\ref{Algebra2}) has a $\nu$-dimensional matrix representation
\begin{eqnarray}
\label{algebrarep finit}
D_{k}=q_{k} \sum_{i=0}^{\nu-1}|i\rangle\langle k-1 | \quad \mbox{for $k=1,\cdots,\nu$}\, , \nonumber \\
E=\sum_{i=0}^{\nu-1}p_{i+1} |i\rangle\langle i | \,.
\end{eqnarray}
Using the standard MPF we have found that the steady-state of the systems with the following dynamical rules
\begin{equation}
\label{dynamic koli}
\begin{array}{lll}
I0 \rightarrow 0I \quad &\mbox{with rate $\Gamma^{I,0}_{0,I}$}&\, ,\cr
I0 \rightarrow KI \quad &\mbox{with rate $\Gamma^{I,0}_{K,I}$}&\, ,\cr
IJ \rightarrow KJ \quad &\mbox{with rate $\Gamma^{I,J}_{K,J}$}&\, ,\cr
IJ \rightarrow 0J \quad &\mbox{with rate $\Gamma^{I,J}_{0,J}$}&\, ,\cr
IJ \rightarrow J0 \quad &\mbox{with rate $\Gamma^{I,J}_{J,0}$}&\cr
\end{array}
\end{equation}
in which $I,J,K\in\{1,\cdots,\nu\}$, can be described by~(\ref{Algebra2}) provided that some constraints are fulfilled.
These constraints can be obtained as follows: according to the MPF the steady-state probability distribution of the system which satisfies
\begin{equation}
\label{ss dist}
H \vert P(\infty) \rangle=0\;,
\end{equation}
is written as
\begin{equation}
\label{steady-state}
|P(\infty)\rangle \propto \Tr[\mathcal{D}^{\otimes L}]\;,
\end{equation}
in which the column vector $\mathcal{D}$ which is defined as
$$
\mathcal{D}=\pmatrix{E\cr D_1 \cr \vdots \cr  D_{\nu}}\;,
$$ 
should satisfy~\cite{BlytheEvans}
\begin{equation}
\label{MPF}
H [\mathcal{D}\otimes\mathcal{D}]=X\otimes \mathcal{D}-\mathcal{D}\otimes X\;.
\end{equation}
This defines the quadratic algebra of the system. The column vector $X$ is a $(\nu+1) \times 1$ auxiliary vector which is assumed to be 
\begin{equation*}
X=\pmatrix{1\cr 0 \cr \vdots \cr  0}\;.
\end{equation*}
Using~(\ref{MPF}), the stochastic Hamiltonian associated with the dynamical rules~(\ref{dynamic koli}) generates a quadratic algebra. 
Requiring this quadratic algebra to be equal to the quadratic algebra introduced in~(\ref{Algebra2}), one finds that 
$\nu(\nu+1)^2$ transition rates in~(\ref{dynamic koli}) besides the parameters $p_k$'s and $q_k$'s should satisfy the following relations
\begin{eqnarray}
\label{Constraint}
\frac{\sum_{K\neq I}^{\nu}\left(\tilde{\Gamma}^{I,J}_{K,J}-\tilde{\Gamma}^{K,J}_{I,J}\right)+\left(\tilde{\Gamma}^{I,J}_{J,0}+
\tilde{\Gamma}^{I,J}_{0,J}\right)}{\tilde{\Gamma}^{J,0}_{I,J}}=\cr
\frac{\sum_{K=1}^{\nu}\left(\tilde{\Gamma}^{K,J}_{J,0}+\tilde{\Gamma}^{K,J}_{0,J}\right)}{\sum_{K=1}^{\nu}\tilde{\Gamma}^{J,0}_{K,J}}=\cr
\frac{1-\sum_{K=1}^{\nu}\tilde{\Gamma}^{K,J}_{0,J}}{\tilde{\Gamma}^{J,0}_{0,J}}=p_J \quad \mbox{for} \quad I,J=1,\cdots,\nu
\end{eqnarray}
in which we have introduced
\begin{equation}
\begin{array}{ll}
\label{newparameter}
\tilde{\Gamma}^{I,J}_{K,J}=q_I\Gamma^{I,J}_{K,J}\; , &  \tilde{\Gamma}^{I,J}_{0,J}=q_I\Gamma^{I,J}_{0,J}\; ,\\
\tilde{\Gamma}^{I,J}_{J,0}=q_I\Gamma^{I,J}_{J,0}\; , & \tilde{\Gamma}^{I,0}_{K,I}=\Gamma^{I,0}_{K,I}\; ,\\
\tilde{\Gamma}^{I,0}_{0,I}=\Gamma^{I,0}_{0,I}\; .
\end{array}
\end{equation}
The parameters $p_k$'s and $q_k$'s in~(\ref{Algebra2}) should be calculated from (\ref{Constraint}) in a consistent way. Here we should solve a set of $\nu\left(\nu+2\right)$ equations with $2\nu$ unknowns i.e. $p_k$'s and $q_k$'s. The number of equations are greater than or equal to the number of the unknowns, hence these set of equations can only be solved under some constraints on the transitions rates in (\ref{dynamic koli}).

In section $3$ these constraints are calculated for $\nu=2$. The transition rates in (\ref{dynamic koli}) should be chosen on the conditions that the
constraints~(\ref{Constraint}) are satisfied thus we can introduce a family of systems which can be described by the generalized quadratic algebra (\ref{Algebra2}).

In the next section we will show that the steady-state probability distribution function of this family of systems are multiplicative.
\begin{figure}[t]
\centering
\includegraphics{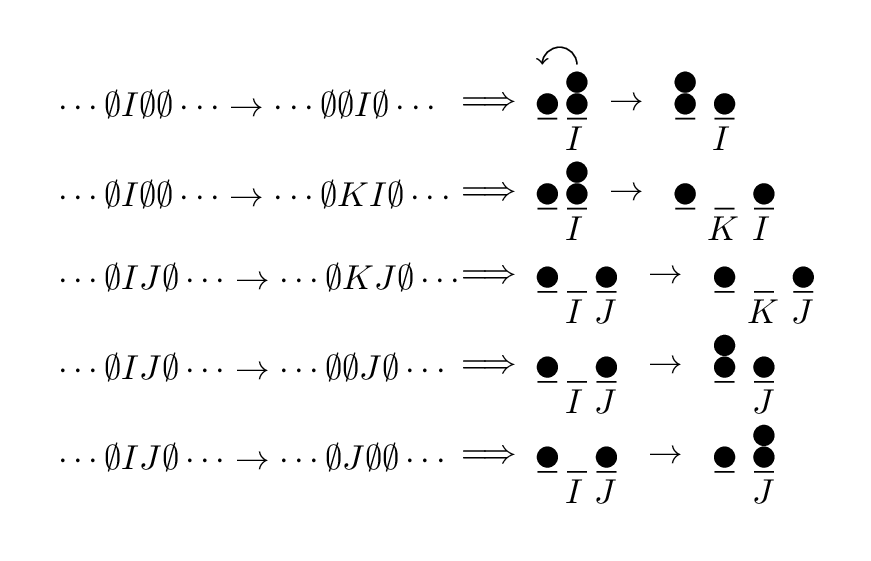}
\caption{\label{fig} Mapping the exclusion process~(\ref{dynamic koli}) onto the ZRP with $I, J, K=1,\cdots,\nu$. In the exclusion process (left) 
a particle is denoted by $I$, $J$ or $K$. An empty lattice-site is also denoted by $\emptyset$. 
A particle can diffuse, leave the system or encounter a type conversion according to~(\ref{dynamic koli}). A particle can 
also be created at an empty lattice-site. In the ZRP (right) a particle (black ball) either hops from one box of type $I$, $J$ or $K$ to another 
box or it can be converted to a new box. Box types can also be changed. }
\end{figure}
\subsection{Relation to the Zero-Range Process}
It is easy to see that the steady-state probability distribution function of the systems that can be described by (\ref{Algebra2}) are product measures.
Let us assume that the total number of particles is denoted by $M$. If the number of empty lattice-sites in front of the $k$'th particle ($k=1,\cdots,M$)
is denoted by $n_{k}$, the steady-state probability of finding the system in a generic configuration
$\{\tau_k n_k\}=\{\tau_1 n_1,\tau_2 n_2,\cdots,\tau_M n_M\}$ is given by
\begin{equation}
\label{PDF}
P({\tau_1 n_1,\tau_2 n_2,\cdots,\tau_M n_M})=\frac{1}{Z_{L}}Tr(D_{\tau_1}E^{n_1}D_{\tau_2}E^{n_2} \cdots D_{\tau_M} E^{n_M})
\end{equation}
in which $Z_{L}$ is a normalization factor and $N_0=\sum_{k=1}^{M}n_k$ is the total number of empty lattice-sites.
Using the matrix representation (\ref{RepAlgebra}) or (\ref{algebrarep finit}) one finds
\begin{equation}
\label{pro}
P(\tau_1 n_1,\tau_2 n_2,\cdots,\tau_M n_M)=f_{\tau_1}(n_1)f_{\tau_2}(n_2)\cdots f_{\tau_M}(n_M)
\end{equation}
in which
\begin{equation}
\label{factor}
f_{k}(n)=q_kp_{k}^{n} \quad \mbox {for $ n=0,1,2,\cdots,\infty $}.
\end{equation}
The relation with the ZRP is now clear. Our disordered reaction-diffusion system defined in~(\ref{dynamic koli}) can be mapped onto a ZRP by identifying $\nu$ different types of particles as $\nu$ different types of boxes and considering the number of empty lattice-sites in front of the particle of type $k$ as the number of particles in the box of type $k$. Hence we have a ZRP with $N_0$ particles and $M$ boxes. This has been shown schematically in Fig.~\ref{fig}. By starting from the ZRP, as described in the
Fig.~\ref{fig}, one can easily recover the constraints~(\ref{Constraint}) using the pairwise balance condition\cite{SRB}.
\section{Examples}

In this section we study the cases $\nu=1$ and $\nu=2$ in details. We look for those disordered reaction-diffusive systems
whose steady-states in terms of the MPF are given by (\ref{Algebra2}).

\subsection{Case $\nu=1$}
As the most simple example let us first consider the case $\nu=1$. It turns out that the most general reaction-diffusion
system in this case consists of the following dynamical rules
\begin{equation}
\label{MDif1}
\begin{array}{lll}
1 \ 0 \longrightarrow 0 \ 1 \quad\mbox{with rate $\Gamma_{01}^{10}$}\cr
1 \ 0 \longrightarrow 1 \ 1 \quad\mbox{with rate $\Gamma_{11}^{10}$}\cr
1 \ 1 \longrightarrow 1 \ 0 \quad\mbox{with rate $\Gamma_{10}^{11}$}\cr
1 \ 1 \longrightarrow 0 \ 1 \quad\mbox{with rate $\Gamma_{01}^{11}$}
\end{array}
\end{equation}
in which $0$ and $1$ represent an empty lattice-site and a particle respectively.
Using (\ref{Constraint}) we can calculate the unknown parameters q and p
\begin{eqnarray}
\label{unknown 1}
q=\frac{\Gamma^{10}_{11}}{\Gamma^{10}_{01}\left(\Gamma^{11}_{10}+\Gamma^{11}_{01}\right)+\Gamma^{11}_{01}\Gamma^{10}_{11}},\ \cr
p=\frac{\Gamma^{11}_{01}+\Gamma^{11}_{10}}{\Gamma^{10}_{01}\left(\Gamma^{11}_{10}+\Gamma^{11}_{01}\right)+\Gamma^{10}_{11}\Gamma^{11}_{01}}.
\end{eqnarray}
Denoting $D_1=D$ and $p_1=p , q_1=q$ the generalized quadratic algebra of the system can be written as
\begin{equation}
\begin{array}{lll}
\label{Algebra21}
DE&=&pD\cr
{D}^2&=&q{D}\, ,
\end{array}
\end{equation}
which has a one-dimensional matrix representation given by
\begin{equation}
\label{matrixrep1}
D=q \quad , \quad E=p\;.
\end{equation}
The partition function of the system can be calculated easily
\begin{equation}
\label{partition 1}
Z_L=Tr[D+E]^L-Tr[E]^L\;.
\end{equation}
The average density of the particles in the system in the long-time limit is given by
\begin{eqnarray}
\rho&=&\frac{1}{Z_L}Tr[D(D+E)^{L-1}]\;.
\end{eqnarray}
In thermodynamic limit $L\rightarrow \infty$ we find
\begin{equation}
\label{expect1}
\rho=\frac{\Gamma^{10}_{11}}{\Gamma^{11}_{10}+\Gamma^{11}_{01}+\Gamma^{10}_{11}}\;.
\end{equation}
It is clear that no phase transitions can occur in this system.

\subsection{Case $\nu=2$}

The most general disordered reaction-diffusion system in the case $\nu=2$ consists of the following dynamical rules
\begin{equation}
\label{Model2}
\begin{array}{lll}
1 \ 0 \longrightarrow 0 \ 1 \quad \mbox {with rate ${\Gamma}_{01}^{10}$}\;, \quad 1 \ 0 \longrightarrow 1 \ 1 \quad \mbox {with rate ${\Gamma}_{11}^{10}$}\;, \cr 1 \ 0 \longrightarrow 2 \ 1\quad \mbox {with rate $\Gamma_{21}^{10}$}\;, \quad 1 \ 1\longrightarrow 1 \ 0 \quad \mbox{with rate $\Gamma_{10}^{11}$}\;, \cr 1 \ 1\longrightarrow 2 \ 1\quad \mbox{with rate $\Gamma_{21}^{11}$}\;, \quad 1 \ 1 \longrightarrow 0 \ 1 \quad \mbox {with rate ${\Gamma}_{01}^{11}$}\;, \cr
1 \ 2\longrightarrow 2 \ 0 \quad \mbox{with rate $\Gamma_{20}^{12}$}\;, \quad 1 \ 2\longrightarrow 2 \ 2\quad \mbox{with rate $\Gamma_{22}^{12}$}\;, \cr 1 \ 2 \longrightarrow 0 \ 2 \quad \mbox {with rate ${\Gamma}_{02}^{12}$}\;, \quad
2 \ 1\longrightarrow 1 \ 0 \quad \mbox{with rate $\Gamma_{10}^{21}$}\;, \cr 2 \ 1\longrightarrow 0 \ 1 \quad \mbox{with rate $\Gamma_{01}^{21}$}\;, \quad 2 \ 1\longrightarrow 1 \ 1 \quad \mbox{with rate $\Gamma_{11}^{21}$}\;, \cr 2 \ 2\longrightarrow 1 \ 2 \quad \mbox{with rate $\Gamma_{12}^{22}$}\;, \quad 2 \ 2\longrightarrow 0 \ 2 \quad \mbox{with rate $\Gamma_{02}^{22}$} \;,\cr
2 \ 2\longrightarrow 2 \ 0 \quad \mbox{with rate $\Gamma_{20}^{22}$}\;, \quad 2 \ 0\longrightarrow 1 \ 2 \quad \mbox{with rate $\Gamma_{12}^{20}$}\;, \cr 2 \ 0\longrightarrow 0 \ 2 \quad \mbox{with rate
$\Gamma_{02}^{20}$}\;, \quad 2 \ 0\longrightarrow 2 \ 2 \quad \mbox{with rate $\Gamma_{22}^{20}$}
\end{array}
\end{equation}
in which the particles of different types are represented by $1$ and $2$ and an empty lattice-site is represented by $0$.
The constraints~(\ref{Constraint}) now reduce to the following equations for 18 transition rates and four unknown parameters $p_{1,2}$ and $q_{1,2}$
\begin{equation}
\label{equation}
\begin{array}{l}
\label{gg}
p_1\tilde{\Gamma}_{01}^{10}=1-\tilde{\Gamma}^{11}_{01}-\tilde{\Gamma}^{21}_{01}\, ,\\
p_1\tilde{\Gamma}_{11}^{10}=\left(\tilde{\Gamma}_{21}^{11}+\tilde{\Gamma}_{01}^{11}+\tilde{\Gamma}_{10}^{11}\right)-\tilde{\Gamma}_{11}^{21}\, ,\\
p_1\tilde{\Gamma}_{21}^{10}=\left(\tilde{\Gamma}_{11}^{21}+\tilde{\Gamma}_{01}^{21}+\tilde{\Gamma}_{10}^{21}\right)-\tilde{\Gamma}_{21}^{11}\, ,\\
p_1\left(\tilde{\Gamma}_{21}^{10}+\tilde{\Gamma}_{11}^{10}+\tilde{\Gamma}_{01}^{10}\right)=\tilde{\Gamma}_{10}^{11}+\tilde{\Gamma}_{10}^{21}+1\, ,\\
p_2\tilde{\Gamma}_{02}^{20}=1-\tilde{\Gamma}^{12}_{02}-\tilde{\Gamma}^{22}_{02}\, ,\\
p_2\tilde{\Gamma}_{12}^{20}=\left(\tilde{\Gamma}_{22}^{12}+\tilde{\Gamma}_{02}^{12}+\tilde{\Gamma}_{20}^{12}\right)-\tilde{\Gamma}_{12}^{22}\, ,\\
p_2\tilde{\Gamma}_{22}^{20}=\left(\tilde{\Gamma}_{12}^{22}+\tilde{\Gamma}_{02}^{22}+\tilde{\Gamma}_{20}^{22}\right)-\tilde{\Gamma}_{22}^{12}\, ,\\
p_2\left(\tilde{\Gamma}_{22}^{20}+\tilde{\Gamma}_{12}^{20}+\tilde{\Gamma}_{02}^{20}\right)=\tilde{\Gamma}_{20}^{12}+\tilde{\Gamma}_{20}^{22}+1\,.
\end{array}
\end{equation}

The partition function of the system is a function of four parameters $p_{1,2}$ and $q_{1,2}$ where these parameters are functions of transition rates.
The partition function of the system can be calculated as follow
\begin{equation}
\label{partitionf2zarekoli}
 Z_L=Tr[D_1+D_2+E]^L-Tr[E]^L=Tr[C]^L-Tr[E]^L
\end{equation}
where we have defined $C=D_1+D_2+E$ for later convenience. Using a two-dimensional matrix representation of the algebra given in~(\ref{algebrarep finit})
we find
\begin{equation}
\label{Cmatrix}
C =\pmatrix{p_1+q_1&q_2\cr q_1&p_2+q_2}.
\end{equation}
The eigenvalues of $C$ can be easily obtained and are given by
\begin{eqnarray}
\label{eigenvalue}
\lambda_{\pm}&=&\frac{1}{2}[p_1+p_2+q_1+q_2 \\
 &&\pm\sqrt{(p_1+p_2+q_1+q_2)^2-4(p_1p_2+p_2q_1+p_1q_2)}]\;. \nonumber
\end{eqnarray}

The stationary average density of the particles of type $1$ and $2$ which are given by
\begin{eqnarray}
\label{density}
\rho_1=\frac{Tr[D_1C^{L-1}]}{Z_L}\, ,\\
\rho_2=\frac{Tr[D_2C^{L-1}]}{Z_L}
\end{eqnarray}
can be calculated using a $2$-dimensional matrix representation of the quadratic algebra. In the thermodynamic limit $L \to \infty$ these quantities are given by
\begin{equation}
\fl
\begin{array}{ll}
\label{average koli}
\rho_1=  
\frac{p_1p_2-p_{2}^2+p_2q_1+2p_1q_2-p_2q_2+p_2\sqrt{\left(p_1+p_2+q_1+q_2\right)^2-4\left(p_2q_1+p_1p_2+p_1q_2\right)}}{2\left(p_2q_1+p_1p_2+p_1q_2\right)\sqrt{(p_1+p_2+q_1+q_2)^2-4\left(p_2q_1+p_1p_2+p_1q_2\right)}}&\, ,\cr \cr
\rho_2= 
\frac{p_1p_2-p_{1}^2-p_1q_1+2p_2q_1+p_1q_2+p_1\sqrt{\left(p_1+p_2+q_1+q_2\right)^2-4\left(p_2q_1+p_1p_2+p_1q_2\right)}}{2\left(p_2q_1+p_1p_2+p_1q_2\right)\sqrt{(p_1+p_2+q_1+q_2)^2-4\left(p_2q_1+p_1p_2+p_1q_2\right)}} &\, . \nonumber
\end{array}
\end{equation}

A disordered reaction-diffusion system with the dynamical rules~(\ref{dynamic koli}) is the most general one whose steady-state can be described by
the generalized quadratic algebra~(\ref{Algebra2}) in the case $\nu=2$. In what follows we will consider two special examples which have already been studied in the literature.

\subsubsection{Special example $1$}
In \cite{BasuMohanty1} the authors have introduced a non-conserved totally asymmetric exclusion process on a ring with internal degrees of freedom where the
transitions allowed are
\begin{equation}
\label{dynamic basu}
\begin{array}{lll}
10 \quad \rightarrow 01 \quad \mbox{with rate $\alpha_{1}=\Gamma_{01}^{10}$} \cr
20 \quad \rightarrow 02 \quad \mbox{with rate  $\alpha_{2}=\Gamma_{02}^{20}$} \cr
11 \quad \rightarrow 21 \quad \mbox{with rate $\beta_2=\Gamma_{21}^{11}$} \cr
12 \quad \rightarrow 22 \quad \mbox{with rate $ \beta_2=\Gamma_{22}^{12}$} \cr
22 \quad \rightarrow 12 \quad \mbox{with rate $\beta_1=\Gamma_{12}^{22}$} \cr
21 \quad \rightarrow 11 \quad \mbox{with rate $\beta_1=\Gamma_{11}^{21}$}\, .
\end{array}
\end{equation}
These rules can be summarized as follow
\begin{equation}
\label{dynamic koli basu}
\begin{array}{l}
I0 \rightarrow 0I \quad \mbox{with the rate $\Gamma^{I0}_{0I}$}\, , \\
IJ \rightarrow KJ \quad \mbox{with the rate $\Gamma^{IJ}_{KJ}$}
\end{array}
\end{equation}
in which $I,J,K\in\{1,2\}$. The system described by the dynamical rules in~(\ref{dynamic koli basu}) is a special case of~(\ref{dynamic koli}).
It should be noted that by defining $p_{1,2}=1/ \alpha_{1,2}$ and $q_2=(q_1\beta_1)/\beta_{2}$
the transition rates in~(\ref{dynamic basu}) satisfy the equations~(\ref{equation}) without exerting any new constraints.
Note that $q_1$ is a free variable and we can choose $q_1=1$ without loss of generality.

We should emphasize that although the quadratic algebra introduced in \cite{BasuMohanty1} is different from our generalized quadratic algebra
which in this case is given by
\begin{equation}
\label{algebra basu}
\begin{array}{ll}
D_kE=p_kD_k \quad &\mbox{for $k\in \{1,2\}$} \, ,\\
D_kD_{k'}=q_kD_{k'}  \quad &\mbox{for $k,k'\in\{1,2\}$}
\end{array}
\end{equation}
they have the same matrix representations.

As we mentioned, all the members of this family of exactly solvable systems can be mapped onto a ZRP. It is not surprising that the authors in~\cite{BasuMohanty1} have
shown that their model can be mapped onto a ZRP.
\subsubsection{Special example $2$}
Another member of the family of models in~(\ref{Model2}) is introduced and studied in ~\cite{jafarpour zeraati}. The transition rates are chosen as follows
\begin{equation}
\label{Model2 spec}
\begin{array}{lll}
1 \ 0 \longrightarrow 0 \ 1 \quad \mbox {with rate $\alpha_1$} \quad \;\;\; 1 \ 0 \longrightarrow 1 \ 1 \quad \mbox {with rate $1$} \cr 1 \ 1\longrightarrow 1 \ 0 \quad \mbox{with rate $1$} \quad \; \; \quad 1 \ 0 \longrightarrow 2 \ 1\quad \mbox {with rate $\beta_1$} \cr 1 \ 1\longrightarrow 2 \ 1\quad \mbox{with rate $\beta_1$}\quad \quad 1 \ 2\longrightarrow 2 \ 0 \quad \mbox{with rate $\alpha_1$} \cr 1 \ 2\longrightarrow 2 \ 2\quad \mbox{with rate $\beta_1$} \quad \quad 2 \ 1\longrightarrow 1 \ 0 \quad \mbox{with rate $\beta_2$} \cr
2 \ 1\longrightarrow 1 \ 1 \quad \mbox{with rate $\beta_2$} \quad \quad 2 \ 2\longrightarrow 1 \ 2 \quad \mbox{with rate $\beta_2$} \cr
2 \ 2\longrightarrow 2 \ 0 \quad \mbox{with rate $\alpha$} \quad \; \,\quad 2 \ 0\longrightarrow 1 \ 2 \quad \mbox{with rate $\alpha_2$} \cr 2 \ 0\longrightarrow 0 \ 2 \quad \mbox{with rate $\alpha_2$} \quad \quad 2 \ 0\longrightarrow 2 \ 2 \quad \mbox{with rate $\beta$}
\end{array}
\end{equation}
in which $\beta=\frac{\beta_1}{\beta_2}$ and $\alpha=\frac{\alpha_1}{\alpha_2}$. The dynamical rules in~(\ref{Model2 spec}) can be summarized as
\begin{equation}
\label{sumarized dynamic}
\begin{array}{lll}
I0 \rightarrow KI \quad &\mbox{with rate $\Gamma^{I0}_{KI}$}\, ,&\cr
IJ \rightarrow KJ \quad &\mbox{with rate $\Gamma^{IK}_{KJ}$}\, ,&\cr
IJ \rightarrow J0 \quad &\mbox{with rate $\Gamma^{IJ}_{J0}$}&
\end{array}
\end{equation}
in which $I,J,K\in\{1,2\}$. We should emphasize that the transition rates in~(\ref{sumarized dynamic}) have to satisfy~(\ref{equation}). This can be achieved by choosing $p_{1,2}=1/ \alpha_{1,2}$ and $q_2=q_1\beta$. This will not exert any new constraints on the transition rates. Note that $q_1$ is a free variable and as the previous case we can choose $q_1=1$ without loss of generality. Analyzing~(\ref{average koli}) reveals that the disordered reaction-diffusion system (\ref{Model2 spec}) has a critical point at $\beta=0$ and $\alpha=2$ which recovers the results already obtained in~\cite{jafarpour zeraati}.

\section{Concluding remarks}

In this paper we have introduced a new generalized quadratic algebra which allows us to study the steady-state of a family of reaction-diffusion 
systems with $\nu$ different species of particles using a matrix method. The number of particles in each species is not a conserved quantity. 
The number of transition rates for a family consisting of $\nu$ different species  is $\nu(\nu+1)^2$. The transition rates are not free  
and have to satisfy some specific constraints. These constraints should be calculated for each system separately.  

For $\nu=1$ the transition rates are given in (\ref{MDif1}). There is no constraints on the transition rates in this case. For the case $\nu=2$ the transition rates in (\ref{Model2}) are not free and besides the parameters $p_{1,2}$ and $q_{1,2}$ have to satisfy the relations in (\ref{Constraint}).
Two special example are studied in detail. One example is a totally asymmetric simple exclusion process with $\nu$ non-conserved internal degrees of freedom, which has
already been introduced and studied in \cite{BasuMohanty1}. 
Another member of this family has been studied recently in~\cite{jafarpour zeraati}. Our results coincide with those obtained in ~\cite{jafarpour zeraati}.



\section*{References}
\begin{thebibliography}{99}
\bibitem{schutz} G. M. Sch\"{u}tz Phase Transitions and Critical Phenomena,vol19, C.Domb and J.Lebowitz eds (Academic, London, 2001).
\bibitem{Benjamini} I. Benjamini, P. A. Ferrari and C. Landim Stoch, Proc. Apple. {\bf 61} 181 (1996).
\bibitem{EvansEurophys} M. R. Evans, Europhys. Lett. {\bf 36} 13 (1996).
\bibitem{BasuMohanty1} U. Basu, P. K. Mohanty, Phys. Rev. {\bf E 82} 041117 (2010).
\bibitem{Vahid} V. Karimipour, Phys. Rev. {\bf E 59}, 205 (1999).
\bibitem{JafarpourRing} F. H. Jafarpour, Physica A, 303 pp.144 (2002).
\bibitem{DerridaEvansHakimPasquier} B. Derrida, M. R. Evans, V. Hakim and V. Pasquier, J. Phys. {\bf A}: Math. Gen.{\bf 26} 1493 (1993).
\bibitem{IPR} A. P. Isaev, P. N. Pyatov, V. Rittenberg, J. Phys. A: Math. Theor. {\bf 34} 5815 (2001).
\bibitem{Alcaraz1} F. C. Alcaraz and V. Rittenberg, Phys. Lett. B. 314 377 (1993).
\bibitem{Alcaraz2} F. C. Alcaraz, S. Dasmahapatra and V. Rittenberg, J. Phys. A: Math. Gen. 31, 845 (1998).
\bibitem{Arndt} P. F. Arndt, T. Heinzel and V. Rittenberg, J. Phys. {\bf A}: Math. Gen. 31 833 (1998).
\bibitem{BlytheEvans} R. A. Blythe, M. R. Evans, J. Phys. {\bf A}: Math. Theor. {\bf 40} R333 (2007).
\bibitem{Andreas} K. Klauck and  A. Schadschneider, Physica A, 271 pp.102 (1999).
\bibitem{EFM} M.R. Evans, P. Ferrari and K. Mallick, J. Stat. Phys. 135, 217 (2009).
\bibitem{PEM} S. Prolhac, M. R. Evans, K. Mallick, J. Phys. A: Math. Theor. 42 165004 (2009).
\bibitem{EvansHanney} M. R. Evans and T. Hanney, J. Phys. A: Math. Gen. {\bf 38} R195 (2005).
\bibitem{EvansBraz} M. R. Evans, Braz. J. Phys. {\bf 30} 42 (2000).
\bibitem{Jafarpour} F. H. Jafarpour, Phys. Rev. {\bf E 83} 041112 (2011).
\bibitem{BasuMohanty} U. Basu, P. K. Mohanty, J. Stat. Mech.  L03006 (2010).
\bibitem{SRB} G. M. Sch\"utz, R. Ramaswamy and M. Barma, J. Phys. {\bf A}: Math. Gen. 29 837(1996).
\bibitem{jafarpour zeraati} S. Zeraati, F. H. Jafarpour, and H. Hinrichsen, Phys. Rev. E {\bf  87}, 062120 (2013).
\end{thebibliography}
\end{document}